\input{aipcheck.tex}

\edef\optionlist{%
   \variorefoptionifavailable        
   draft,%
   \psnfssproblemoption              
   tnotealph}

\documentclass[\optionlist,sort]{aipproc}

   \usepackage[draft=false,bookmarksopen]
              {hyperref}
\layoutstyle{6x9}

\received{x}
\accepted{x}

\usepackage{ttct0001}

\usepackage{shortvrb}
\MakeShortVerb\|

\newcommand{\etal}{{et al.}}

\newcommand{\Msun}{\>{\rm M_{\odot}}}

\begin{document}

\author{T. Bogdanovi\'c}{
  address={Department of Astronomy \& Astrophysics, The Pennsylvania State University},
  altaddress={Center for Gravitational Wave Physics, Institute for Gravitational Physics and Geometry}, 
  email={tamarab@astro.psu.edu},
}

\author{B. D. Smith}{
  address={Department of Astronomy \& Astrophysics, The Pennsylvania State University},
  email={britton@astro.psu.edu},
}

\author{M. Eracleous}{
  address={Department of Astronomy \& Astrophysics, The Pennsylvania State University},
  altaddress={Center for Gravitational Wave Physics, Institute for Gravitational Physics and Geometry},   
  email={mce@astro.psu.edu},
}

\author{S. Sigurdsson}{
  address={Department of Astronomy \& Astrophysics, The Pennsylvania State University},
  altaddress={Center for Gravitational Wave Physics, Institute for Gravitational Physics and Geometry},  
  email={steinn@astro.psu.edu},
}

\title{Electromagnetic Signatures of Massive Black Hole Binaries}
\date{}

\keywords{supermassive black hole binaries, electromagnetic signatures, SPH simulations}
\classification{98.62.Js}

\begin{abstract}
We model the electromagnetic emission signatures of massive black hole
binaries (MBHBs) with an associated gas component. The method
comprises numerical simulations of relativistic binaries and gas
coupled with calculations of the physical properties of the emitting
gas. We calculate the accretion powered UV/X-ray and H$\alpha$ light
curves and the H$\alpha$ emission line profiles. The simulations have
been carried out with a modified version of the parallel tree SPH code
{\it Gadget}. The heating, cooling, and radiative processes for the
solar metallicity gas have been calculated with the photoionization
code {\it Cloudy}. We investigate gravitationally bound, sub-parsec
binaries which have not yet entered the gravitational radiation
phase. The results from the first set of calculations, carried out for
a coplanar binary and gas disk, suggest that the outbursts in the
X-ray light curve are pronounced during pericentric passages and can
serve as a fingerprint for this type of binaries if periodic outbursts
are a long lived signature of the binary. The H$\alpha$ emission-line
profiles also offer strong indications of a binary presence and may be
used as a criterion for selection of MBHB candidates for further
monitoring from existing archival data. The orbital period and mass
ratio of a binary could be determined from the H$\alpha$ light curves
and profiles of carefully monitored candidates. Although systems with
the orbital periods studied here are not within the frequency band of
the {\it Laser Interferometer Space Antenna} ({\it LISA}), their
discovery is important for understanding of the merger rates of MBHBs
and the evolution of such binaries through the last parsec and towards
the detectable gravitational wave window.
\end{abstract}

\maketitle


\bigskip

\section{Introduction}

Massive black hole binaries (MBHBs) are expected to be the most
luminous sources of gravitational radiation in the universe. The
simultaneous identification of the gravitational radiation and
electromagnetic counterpart from MBHBs would allow unprecedented tests
of the physics of massive black holes, such as accretion, and also
offer an alternative method to constrain cosmological
parameters. However, there are only a handful of MBHB candidates known
from observations so far and none of them are expected to merge within
our lifetimes. Although galaxy mergers are the natural places to look
for MBHBs \citep*{bbr,valtaoja89,milosavljevic01,yu02}, not every such
galaxy forms a MBHB which will merge within the Hubble time, hence the
difficulty in determining the merger rates. Practical obstacles in the
direct identification arise for several reasons. Firstly, fairly high
spatial resolution and accuracy in position measurements are required
to resolve a binary with two active nuclei. For example, the spatial
resolution required to resolve an intermediate binary with orbital
separation of 1pc at the distance of $\sim$100$\,$Mpc is about
2$\,$mas. With spatial resolutions currently achievable, it is easier
to spot and resolve wide binaries, like NGC 6240, whereas it has been
suggested that MBHBs spend a major fraction of their life time as
intermediate, hard binaries \citep{bbr}.  It follows that a better
understanding of electromagnetic signatures of galactic nuclei hosting
the massive binaries is essential in order to recognize the presence
of MBHBs and put tighter constraints on their merger rates.

\section{Numerical simulations}

We have carried out smoothed particle hydrodynamical (SPH) simulations
of the binary and gaseous component, and we have characterized the
physical properties of the gas by calculating heating, cooling, and
radiative processes as an integral part of simulations. Based on these
results we have calculated the accretion-powered continuum and
H$\alpha$ light curves, as well as the H$\alpha$ emission line
profiles emerging from the inner parts of a gas disk on a scale of
$10^{-4}-1$ pc.

We have used {\it Gadget} \citep*{springel01,springel05} to carry out
the MBHB simulations. {\it Gadget} is a code for collisionless and
gas-dynamical cosmological simulations. It evolves self-gravitating
collisionless fluids with a tree $N$-body approach, and collisional
gas by SPH.  We have performed several modifications to the code in
order to treat the two massive black holes relativistically. We have
introduced the black holes as collisionless massive particles with
pseudo-Newtonian Paczynsky-Wiita potentials \citep{paczynsky}.  The
calculations account for the decay of a black hole binary orbit
through emission of gravitational radiation. We have estimated the
accretion rate of gas and resulting accretion luminosity. We assume
that sources of illumination are powered by accretion onto the massive
black holes. After accretion on either of the two black holes becomes
significant, UV and soft X-ray radiation emitted from the innermost
accretion flow photo-ionizes the gas.

The approximate method was developed for calculation of heating and
cooling of a gas with metals. We have constructed a number of cooling
maps using the photoionization code {\it Cloudy}
\citep{ferland}.  The parameter grid is read in by the simulation code, and
radiative heating and cooling rates are linearly interpolated from the
existing grid points. The cooling maps are calculated in the parameter
space of {\it density} and {\it temperature} of the gas and {\it
intensity} of ionizing radiation. The range of parameter values for
which the maps were computed is as follows: $10^9\,{\rm cm^{-3}} < n <
10^{19}\,{\rm cm^{-3}}$, $2000\,{\rm K} < T < 10^8$ K, and $0\; {\rm
erg \;cm^{-2} \, s^{-1}} < J < 10^{17}\, {\rm erg \;cm^{-2} \,
s^{-1}}$. Because one of the assumptions in {\it Cloudy} is that the
electrons are non-relativistic, the present range of its validity
extends to temperatures below roughly $10^9$ K. Gas at higher
temperatures than this is commonly encountered in our simulations; in
such cases we calculate heating and cooling rates by linearly
extrapolating the grid values. In addition to the cooling processes 
included in {\it Cloudy} we also consider Compton cooling from a
thermal distribution of non-relativistic electrons, which may become a
significant coolant for the hottest and densest gas in the nuclear
region.  Since {\it Cloudy} includes a detailed treatment of radiation
processes, we have also used it to calculate the H$\alpha$ line
intensity and optical depth, the electron scattering optical depth,
and the neutral hydrogen column density for each gas cell. The 
H$\alpha$ emission line profiles have been calculated taking into
account the relativistic Doppler shift and the gravitational redshift in the 
potential well of a Schwarzschild black hole. We compute line profiles 
emerging from the gaseous disk under assumption that the observer 
is located at a distance $d\to \infty$, at $i=30^{\circ}$ to the axis 
orthogonal to the plane of the binary orbit. 

The masses of the primary and secondary black holes are $10^{7}\Msun$
and $10^{8}\Msun$. At the beginning of the simulation the black holes
start from the apocenters of their orbits. The initial values of the
eccentricity and semi-major axis are 0.7 and 3007, respectively, with
the semi-major axis in units of $r_g$ (used as the unit of distance
throughout this paper), where $r_{g}\equiv
GM_{BH}/c^{2}=1.48\times10^{13}\,M_{8}\;{\rm cm}$, and $M_{BH}=10^8
M_8\,\Msun$ is the mass of a black hole. The period of the binary,
calculated at the beginning of the simulation is 15.67 yr. The
accretion disk is initially associated with the primary black hole. It
is co-planar with the binary orbit and it extends from $\xi_{in}=100$
to $\xi_{out}=2000$ ($\xi_{out}$ translates to $0.01$ pc in physical
units, a size expected for AGN accretion disks). The mass of the disk
is $10^{4}\Msun$ and the number of gas particles in the disk is
$10^5$. The initial surface density distribution in the disk is chosen
to be $\Sigma\propto\xi^{-1.5}$. The initial temperature of the disk
is $T=2000$ K. The numerical viscosity parameter ($\alpha_G =
10^{-6}$), accretion efficiency ($\eta = 10^{-2}$), and metallicity of
the gas ($\log\, Z = 0$) are kept constant. For detailed description
of numerical methods used see
\citet{bogdanovic06}.

\section{Results}

\begin{figure}[h]
\centerline{
\resizebox{2.5in}{!}{\includegraphics{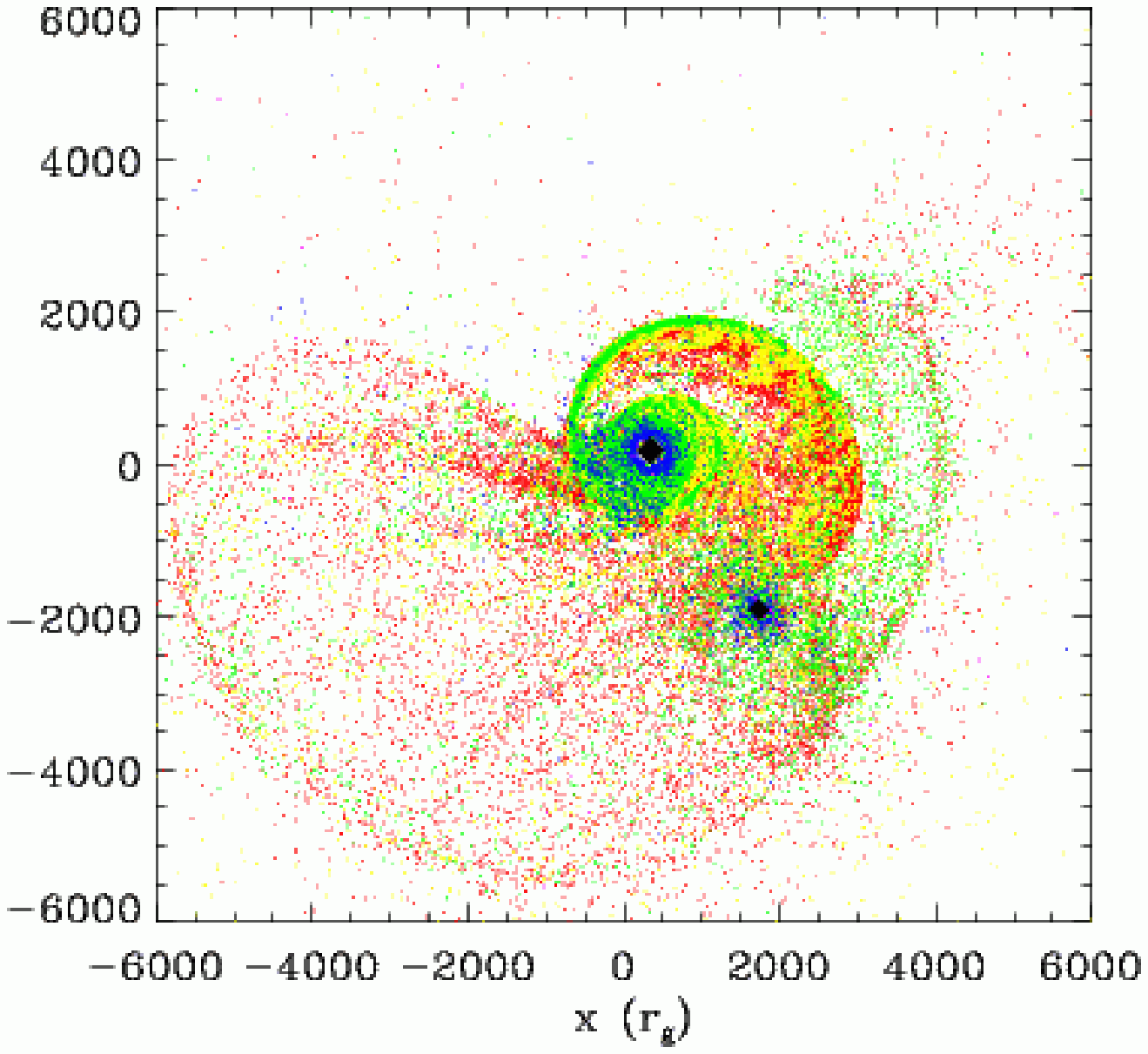}}
\resizebox{2.5in}{!}{\includegraphics{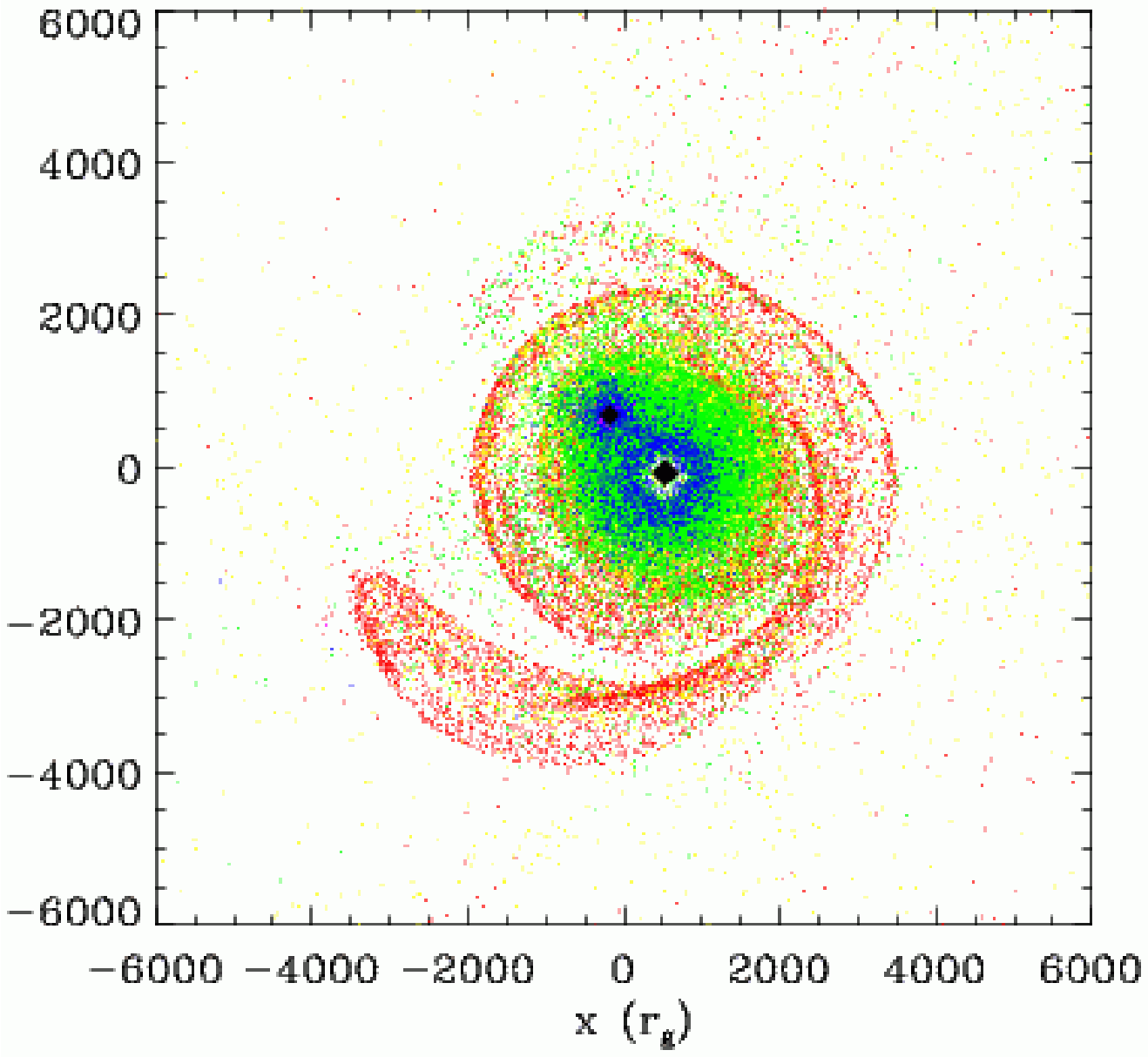}}}
\caption{The snapshots from the simulation showing the binary and gas 
(projected to the plane of the binary orbit) at 9.4 and 22.8 years,
left and right panel, respectively. The rotation of the binary and the
disk is counter-clockwise. The temperatures of gas particles are
marked with color: red $T < 10^4$ K; yellow $10^4 {\rm K} < T< 10^6$
K; green $10^6 {\rm K}< T < 10^8$ K; blue $10^8 {\rm K} < T < 10^{10}$
K; and violet $T > 10^{10}$ K. }
\label{fig_xy_sr}
\end{figure}

In the scenario described here the binary is co-rotating with the
accretion disk and the gas is assumed to have a solar metallicity.
Figure~\ref{fig_xy_sr} shows the morphology and temperature of gas at
two different times in the simulation.
After the passage of the secondary black hole through the disk the
disk density settles in the range $10^{10}-10^{14}\; {\rm
cm^{-3}}$. The disk gas remains well above the Toomre threshold for
gravitational instability. The speed of sound is about $5\,{\rm
km\,s^{-1}}$ in the cold disk and $\sim 10^2\,{\rm km\,s^{-1}}$ in the
hotter gas component. The temperature of the gas reaches the highest
value (T$\sim10^{12}$ K) after the shock is formed by the
secondary. On a time scale of months, the temperature of the gas falls
below $10^{10}$ K due to the combined effects of radiation and
adiabatic expansion. The hot component of the gas spends a significant
amount of time in the temperature range $10^4-10^8$ K. In this regime
the radiative cooling is dominated by free-free emission and
recombination radiation, while inverse Compton emission becomes
inefficient in comparison. On the other hand, the cold component of
the gas, confined to the spiral arm retains a temperature of
$10^3-10^4$ K.
\begin{figure} [h]
\resizebox{3.2in}{!}{\includegraphics{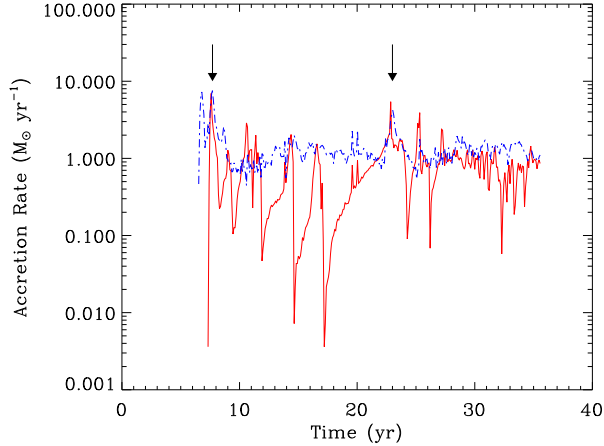}}
\caption{The effective accretion rate on the primary 
({\it solid, red line}) and secondary ({\it dashed, blue line}) black
hole. The accretion rate curves in our calculation can be translated
into UV/X-ray light curves from the emission sources, where
$1\Msun\,{\rm yr^{-1}} \propto 10^{43}\,{\rm erg\,s^{-1}}$ of UV/X-ray
luminosity. The arrows mark the times of pericentric passages of the
binary.}
\label{fig_acc_sr}
\end{figure}
After the first pericentric passage of the secondary the accretion
rate reaches $10 {\rm \Msun\;yr^{-1}}$ (Figure~\ref{fig_acc_sr}),
comparable to the Eddington rate for this system, $\dot{M}_E \approx
30\,M_8\, (\eta/0.01)^{-1}\;{\rm \Msun\;yr^{-1}}$. During the
remainder of the simulation the accretion rate remains at a nearly
constant level of about $1 {\rm \Msun\;yr^{-1}}$. This implies that
the total accretion luminosity is just below the Eddington luminosity
($L_{E}=1.51\times10^{46}M_{8}\,{\rm erg\,s^{-1}}$) for a few years
after the first pericentric passage of the binary and later settles at
luminosity $\sim 10^{-2}\,L_{E}$. During this period the bolometric
luminosity of the disk, powered by shocks and illumination, is found
to be $\sim 10^{45}\,{\rm erg\,s^{-1}}$ on average, with fluctuations
of up to 2 orders of magnitude.  Additionally, the pericentric
passages, in the 7th and 23rd years of the simulation can be easily
discerned in the accretion rate curves of both black holes
(Figure~\ref{fig_acc_sr}).

We calculate the X-ray luminosity contributed by two different
mechanisms: the X-ray emission powered by accretion and that powered
by bremsstrahlung emission from the hot gas. The accretion rates onto
the primary and secondary black holes in the simulation result in a
UV/X-ray luminosity of about $10^{43}~\,{\rm erg\,s^{-1}}$ and up to
$10^{44}~\,{\rm erg\,s^{-1}}$ during times of high accretion rate
(Figure~\ref{fig_acc_sr}). We calculate the thermal bremsstrahlung
emission from the population of relativistic electrons hotter than
$10^7$ K. The electrons are heated in the process of photoionization
of the gas by the accretion powered sources and in shocks taking place
in the gas gravitationally perturbed by the secondary black hole. The
calculation of the size of the emitting volume from a SPH simulation
is complicated by the finite spatial resolution, as set by the
characteristic size assigned to each gas particle (i.e., the smoothing
length, $h_{sml}$). In our calculations $h_{sml} > 10\,r_g \approx
10^{14}\,{\rm cm}$. It follows that as a result of ``smoothing'', the
bremsstrahlung X-ray emission from shocks, which typically occur on
scales smaller than $h_{sml}$ in our simulations, is not resolved. In
order to address this uncertainty we calculate analytically the values
of some physical properties of the X-ray emitting gas following the
analytical treatment of shock structure in \citet{hm79}. We find that
the resulting bremsstrahlung X-ray luminosity from the {\it shocked}
gas is in the range $10^{40}-10^{42}\,{\rm erg\,s^{-1}}$, with the
peaks reaching up to $10^{43}\,{\rm erg\,s^{-1}}$. In the second step
we estimate the bremsstrahlung X-ray luminosity emitted from the {\it
photoionized} portion of the gas. We identify the gas cells
responsible for most of the X-ray emission powered by photoionization
as low density cells with temperatures higher than $T > 10^7\,{\rm
K}$. The number of photoionized gas particles that contribute to the
X-ray bremsstrahlung emission is about a few percent of all gas
particles. However, the large volume of this halo makes up for its
small mass and low number density. Thus, the resulting bremsstrahlung
luminosity from the halo is $10^{41}-10^{42}\,{\rm erg\,s^{-1}}$,
comparable to the estimated contribution of the shocked regions. We
emphasize that the estimates of the bremsstrahlung X-ray luminosity
presented here are based on astrophysically motivated but simplified
assumptions and that they should be regarded as constraints rather
then the exact values.

The bremsstrahlung X-ray luminosity of the gas is, in general, lower
than the accretion powered UV/X-ray luminosity, although during the
times of pericentric passage the peak bremsstrahlung X-ray luminosity
may be comparable to that of the accretion powered X-ray
luminosity. The latter X-ray light curve also exhibits peaks during
pericentric passages of the binary, thus such peaks are good markers
of such events.  The estimated level of the total X-ray emission
(accretion powered plus bremsstrahlung) should be observable to a
redshift of $z \le 2$ during the outburst phases. A calculation
following the X-ray light curve variability over a large number of
orbits is necessary in order to confirm that the periodicity is a long
lived signature of the binary.
\begin{figure} [h]
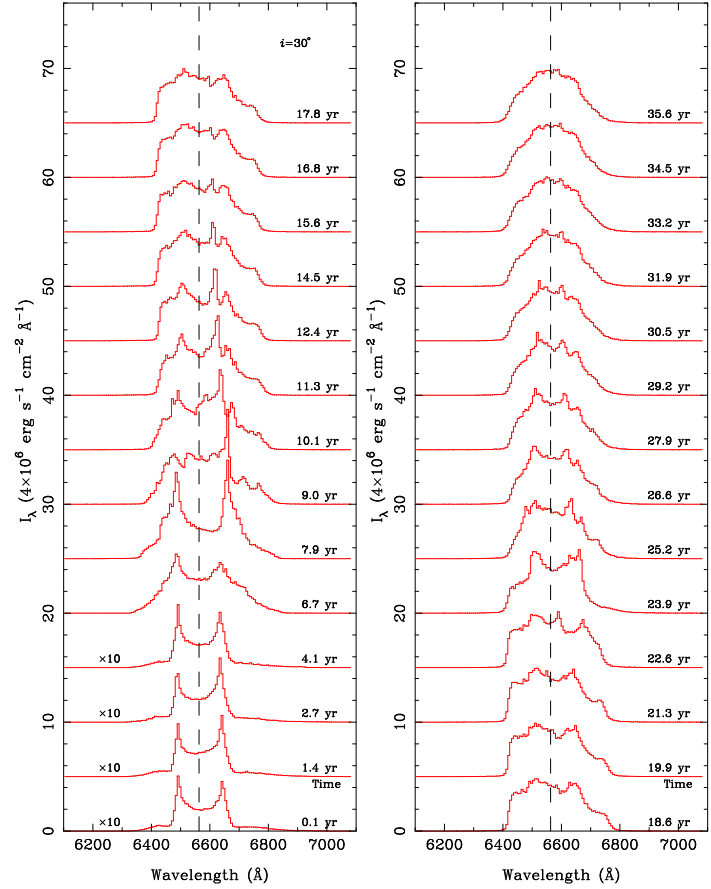

\resizebox{1.8in}{!}{\includegraphics{bogdanovic_fig3a.ps}}
\resizebox{1.8in}{!}{\includegraphics{bogdanovic_fig3b.ps}}
\caption{A sequence of the selected H$\alpha$ emission-line profiles. 
The intrinsic intensity of profiles if plotted against wavelength. The
first 4 profiles in the sequence are multiplied by a factor of 10, so
that they can be represented on the same intensity scale with the
other profiles. The corresponding time from the beginning of the
simulation is plotted next to each profile. The inclination of the
plane of the binary with respect to the observer is as marked on the
figure. The vertical dashed line at 6563~{\rm \AA} marks the H$\alpha$
rest frame wavelength.}
\label{fig_prof_solar}
\end{figure}

By modeling the H$\alpha$ light curves we find that after the
beginning of accretion the H$\alpha$ luminosity gradually reaches
$10^{39}-10^{40}\;{\rm erg\, s^{-1}}$. Sources with such H$\alpha$
luminosities are observable out to the distance of the Virgo Cluster
and possibly up to a distance of 100 Mpc. We also calculate the broad
H$\alpha$ emission-line profiles in order to better illustrate the
kinematics of the gas (see Figure~\ref{fig_prof_solar}).  Initially,
from the unperturbed disk we observe low intensity double-peaked
emission line profiles but they gradually depart from this shape as
the perturbation propagates through the gas. The profiles appear
variable on a time scale of months to years, both in shape and
width. The profiles show an extended, low intensity red wing, most
pronounced during the periods of increased accretion and after the
times of pericentric passages. During such events a number of emitting
gas particles find themselves deeper in the potential well of the
binary. The emitted photons that leave the potential well are
gravitationally redshifted and Doppler boosted, thus contributing to
the extended red wing and the blue shoulder of the relativistic
H$\alpha$ profiles. Because the relativistic effects emphasize the
distinctive features in the H$\alpha$ emission line profiles during
and after the pericentric passages of the binary they are in general
case a helpful marker which can be used to infer the orbital period of
the binary from the observed profile sequences. The trailed
spectrogram (a 2D map of the H$\alpha$ intensity against observed
velocity, not shown here) reveals the repetitive behavior in segments
between 7 to 22 years and 22 to 36 years. The width of the H$\alpha$
profile increases after the pericentric passages of the system,
reflecting the inflow of gas towards the primary. Additionally the
widening of the profile appears asymmetric and shifted towards the red
with respect to the pre-pericentric sequence of profiles. The shift in
the emission line profile sequence is a signature of the motion of the
accretion disk which follows the primary on its orbit. In general, we
expect the signature of the velocity curves of the two black holes to
appear in the emission line profiles at pericentric passages, because
that is when the binary interaction with the gas is most intense. If
such features could be discerned in the observed H$\alpha$
emission-line profiles, they would confirm the existence of the
binary. In practice, one may constrain the period and the mass ratio
of the binary from the periodicity and projected velocity components
of the two black holes, all measured from the H$\alpha$ profile
sequence. The determination of the individual black hole masses is
more challenging, because it requires knowledge of the inclination of
the binary orbit with respect to the observer. Also, before any of the
binary parameters can be measured from observations of light curves
and the H$\alpha$ line sequence, it is necessary to follow a few
revolutions of the binary.

\section{Conclusions}

Since there are reasons to expect that MBHBs spend the largest
fraction of their lives at sub-parsec scales \citep{bbr} this phase
naturally emerges as an optimal one for the study of the observational
signatures of such binaries. Based on the first set of calculations,
one portion of which is presented here, we find that X-ray outbursts
should occur during pericentric passages of a co-planar binary. We
suggest that periodical X-ray outbursts should persist as long as
there is a supply of the nuclear gas that the MBHB can interact
with. A calculation of a much larger number of binary orbits is needed
in order to confirm that the periodic outbursts are a long lived
signature of the binary. The current limitation of the numerical
method used in this study is that the calculation of a large number of
orbits along with the hydrodynamics and radiative heating and cooling
becomes prohibitively expensive.

Except for the recurrent outbursts in the X-ray light curve the
signature of a binary is most easily discernible in the H$\alpha$
emission line {\it profiles}. The irregular shape of the broad
H$\alpha$ emission line profiles can be used as a first set of
criteria when searching for MBHBs. Once candidates are selected from a
large spectroscopic survey (e.g., the SDSS) they can be monitored over
long time intervals to look for the time-dependent signature of a
binary. Based on simulation described here, for a mass ratio of 0.1,
the wavelength shift in the H$\alpha$ emission lines over the course
of an orbital period should be measurable. If one can follow the
regular variations of the line over a few cycles, one could constrain
the properties of the binary. If the signatures of binaries are found
in observations, they could be used to estimate the number of MBHBs in
this evolutionary stage and whether MBHBs indeed evolve quickly
through the last parsec.

\section{Acknowledgments}
T.B. gratefully acknowledges the support of the Center for
Gravitational Wave Physics.


\end{document}